\documentstyle[aps,eqsecnum,psfig]{revtex}
\begin{document}
\draft
\preprint{}                           
\title{A New Approach To Relativistic Gaussian Basis Functions: 
Theory And Applications} 
\author{Rajat K. Chaudhuri\footnote[1]{rajat@iiap.ernet.in}, 
Prafulla K. Panda\footnote[2]{prafulla@iiap.ernet.in} and 
B. P. Das\footnote[3]{das@iiap.ernet.in}}
\address{Non-Accleration Particle Physics Group,\\ Indian Institute of Astrophysics, 
Bangalore-500034, India}
\maketitle
\begin{abstract}
We present a new hybrid method to solve the relativistic Hartree-Fock-Roothan 
equations where the one- and two-electron radial integrals are evaluated 
numerically by defining the basis functions on a grid. This procedure reduces 
the computational costs in the evaluation of two-electron radial integrals. 
The orbitals generated by this method are employed to compute the ionization 
potentials, excitation energies and oscillator strengths of alkali-metal atoms 
and elements of group IIIA through second order many-body perturbation theory 
(MBPT). The computed properties are in excellent agreement with the experiment 
and other correlated theories.
\end{abstract}

\section{Introduction}
The two most critical choices in the application of many-body perturbation 
theory (MBPT) \cite{Freed,Brandow,Lindgren} (relativistic as well as 
non-relativistic) to atomic and molecular systems involve the appropriate 
selection of basis functions and the partitioning of the full Hamiltonian 
$H$ into a zeroth order Hamiltonian $H_0$ and a perturbation $V$ 
\cite{Kelly,Finley,Rajat}. These choices become extremely important when highly 
accurate estimates of various properties are demanded from low order 
perturbative computations. Intensive research has focused on developing 
appropriate basis sets and methods \cite{Finley,Rajat,Malrieu,Mukherjee} to 
minimize the error between the theoretically computed properties and its 
experimental value. The strong dependence of the convergence of MBPT on the
choice of $H_0$ was first demonstrated by Kelly \cite{Kelly} in his pioneering
work on beryllium atom. Using a $V^{N-1}$ instead of the traditional $V^N$ potential for
the excited orbitals, Kelly obtained a vast improvement in the perturbative
convergence for that atom. He also demonstrated that more rapid convergence can be achieved
from a shifted denominator that corresponds to the summation of a certain class of
diagrams to all order.

It is well-known that the theoretical treatment of the heavy atoms must incorporate
certain special features that are not essential for the light atoms. This is 
largely due to the fact that the relativistic effects are so large for heavy atoms 
that it is imperative to treat them by using the relativistic Dirac equation. Despite its 
enormous computational complexity and cost, tremendous progress has been made over the
past decade and a half in solving the four-component Dirac equations for many-electron 
systems using numerical Dirac-Fock (DF) and the finite basis set expansion (FBSE) method 
\cite{Kim,Malli,Ishikawa,Dyall,Grant,Aerts,Clementi,Mohanty,Kello,Sekino,Johnson,Eliav,Oster}. 
The numerical atomic DF self-consistent field (SCF) calculation is more compact and accurate 
but its extension  to molecular systems (multi-center many-electron systems) is cumbersome. Moreover,
the generation of virtual orbital is tedious and frequently encounters convergence
difficulties. The FBSE method, on the other hand, is rather simple
and its extension to molecules is straightforward. Also, the
generation of occupied and virtual orbitals do not require separate computations.

The success of the relativistic FBSE method lies in its proper imposition of 
kinetic-balance condition \cite{Stanton} between the large and small component 
spinor which in essence can be regarded as a proper boundary condition upon the 
basis set. Several papers by Grant {\it et al.} \cite{Grant} and Parpia {\it et al} 
\cite{Parpia} among others document the success of the relativistic FBSE method. 
However, in their finite basis set calculation for light to heavy atoms 
(Z=2-50 and 80), Grant {\it et al.}\cite{Grant} employed kinetically balanced 
Slater-type orbitals (STOs) which have the correct functional behavior but are particularly
unsuitable for analytical self-consistent field (SCF) molecular calculations. Gaussian 
type orbitals (GTOs) or contracted Gaussian type orbitals (CGTO), on the other hand, 
are suitable in the evaluation of multi-center integrals in molecules. It was shown by 
Ishikawa {\it et al.} \cite{Binning} that GTOs can give rise to a natural description 
of the relativistic wave-functions within a finite nucleus.

The most important feature of the FBSE method in STO [$\exp(-\zeta r$)] or GTO 
[$\exp(-\zeta r^2$)] framework is to determine the appropriate exponential 
parameter $\zeta$, because the quality of the wave-function largely depends
upon this parameter and in recent years there has been an increased interest
in finding out the appropriate exponential parameter and contraction coefficients 
(for CGTO) that can provide correct functional behavior of the relativistic 
wave-functions at the nucleus \cite{Clementi,Clementi1,Matsuoka,Hess,Malli1}. For 
instance, Matsuoka {\it et al.} \cite{Matsuoka} have reported accurate configuration 
average DF energies for various atoms through the FBSE method using kinetically 
balanced well-tempered basis set in the framework finite nuclear size approximation. 
While Matsuoka {\it et al.} \cite{Matsuoka} used a well-tempered Gaussian basis set 
in computing DF energies, Clementi {\it et al.} \cite{Clementi} employed kinetically 
balanced geometric-type exponent for the Gaussian primitives and obtained
DF energies for various atoms that are comparable to the numerical DF value 
\cite{Desclaux}. Later Malli {\it et al.} \cite{Malli1} reported 
all-electron {\it ab-initio} fully relativistic DF and DF-Breit calculations for 
polyatomic systems using a relativistic universal Gaussian basis set and recently 
Pernpointer {\it et al.}\cite{Hess} employed a relativistic CGTO basis set in 
their relativistic coupled cluster calculation for the nuclear quadrupole moment of 
CsF. Though the slater (STO) and Gaussian (GTO) types of basis functions are most 
widely used in atomic many-body calculations, this choice is, in principle, arbitrary. 
Since, it is beyond the scope of this present work to discuss this aspect at length, 
we refer the review articles by Grant \cite{Grant1} and Sapirstein \cite{Sapirstein} 
for details.

We have developed a numerical procedure to solve the atomic relativistic DF-SCF 
equations using the FBSE method. This new approach is basically a hybrid of numerical 
and analytical DF (FBSE) methods. Here, the large and small component radial functions 
are expanded in terms of Gaussian primitives on a grid using appropriate constraints 
on the small component radial basis to impose the kinetic-balance condition. While the 
large and small component part of the radial functions are generated (on a grid) 
through the FBSE procedure, the one- and two-electron radial integrals are evaluated 
numerically to avoid the complicated analytical expressions for the two-electron direct 
and exchange radial integrals (the analytical evaluation of one electron radial
integral is rather straight forward). This is the part which differs from the 
conventional FBSE method for solving DF-SCF or HF-SCF equations. This procedure 
(numerical computation of two-electron integrals) also provides an easy way to 
reduces the $N*(N+1)/2$ operations to $N_c$ operations ($N$ and $N_c$ corresponds 
to the number of basis set and occupied orbitals, respectively) in DF-SCF computation, and,
thereby reduces the computational time of relativistic self-consistent field calculations 
for heavy atoms. In the perturbative computations of ground and excited state properties, 
the two-electron radial integrals are also directly computed (numerically) wherever they 
appear to avoid the two-electron integral storage problem. In this paper, we  present some
pilot calculations of the ionization potentials and excitation energies of alkali-metal
and elements of group IIIA computed through second order MBPT using relativistic 
wave-functions obtained from the hybrid DF-SCF approach.

In Sec. II, we describe the hybrid DF-SCF method that has been used to generate the
relativistic single-particle atomic orbitals for post-Dirac-Fock computations. Sec. III 
briefly reviews the background of the MBPT approach for computing ionization potentials
(IP), electron affinities (EA) and excitation energies (EE). The numerical results are 
presented in Sec. IV and compared with other perturbative calculations where available.
We make some concluding remarks in Sec. V.

\section{Hybrid Relativistic Hartree Fock Roothan Equation}
The Dirac-Coulomb Hamiltonian for a many-electron system can be conveniently written as
\begin{equation}
H=\sum_{i=1}^N[c\vec\alpha_i.\vec p_i+(\beta_i-1)mc^2+V_{nuc}(r_i)]+\frac{1}{2}
\sum_{i\ne j} \frac{e^2}{|\vec r_i-\vec r_j|}
\end{equation}
in which the Dirac operators $\vec\alpha$ and $\beta$ are expressed by the matrices
\begin{equation}
\vec \alpha =\left(\begin{array}{lr} 0\hspace{0.2in}\vec\sigma\\\vec\sigma\hspace{0.2in}
0 \end{array}
\right) \hspace{0.3in}
\beta =\left(\begin{array}{lr} I \hspace{0.35in}0\\0\hspace{0.2in}-I\end{array}
\right) \hspace{0.3in}
\end{equation}
where $\bar\sigma$ stands for the Pauli matrices and $I$ is the 2x2 unit matrix.

In the central field approximation, the SCF equations are determined by minimizing
the energy functional $E$ with respect to $\Phi$, where $E$ is given by
\begin{equation}
E=\langle\Phi|\sum_{i=1}^N[c\vec\alpha_i.\vec p_i+(\beta_i-1)mc^2+V_{nuc}(r_i)]+
\frac{1}{2}\sum_{i\ne j}\frac{e^2}{|\vec r_i-\vec r_j|}|\Phi\rangle
\end{equation}
and determinantal wave-function (antisymmetric) $u$ is built from single particle
orbitals
\begin{equation}
u(r,\theta,\phi)=
\left(\begin{array}{c} r^{-1}P_{n\kappa}(r)\chi_{\kappa m}(\theta,\phi)\\
                       ir^{-1}Q_{n\kappa}(r)\chi_{-\kappa m}(\theta,\phi)
\end{array}\right) 
\end{equation}
where $r^{-1}P_{n\kappa}(r)$ and $r^{-1}Q_{n\kappa}(r)$ are the large and small
component radial wave-functions, respectively that satisfy the orthonormality
condition
\begin{equation}
\int_0^{\infty}dr [P_{n\kappa}(r)P_{n^{\prime}\kappa}(r)+
Q_{n\kappa}(r)Q_{n^{\prime}\kappa}(r)]=\delta_{nn^{\prime}}
\end{equation}
Here, the quantum number $\kappa$ classifies the orbital according to their
symmetry and is given by
\begin{equation}
\kappa=-2(j-l)(j+\frac{1}{2})
\end{equation}
where $l$ is the orbital quantum number and $j=l\pm\frac{1}{2}$ is the total 
angular quantum number. Here, the spinors $\chi_{\kappa m}(\theta,\phi)$ are given
\begin{equation}
\chi_{\kappa m}=\sum_{\sigma\pm\frac{1}{2}}C(l\frac{1}{2}j;m-\sigma,\sigma)
Y_{l,m-\sigma}(\theta,\phi)\eta_{\sigma}
\end{equation}
where $C(l\frac{1}{2}j;m-\sigma,\sigma)$ and $Y_{l,m-\sigma}(\theta,\phi)$ represent
the Clebsch-Gordon coefficients and the normalized spherical harmonics, respectively,
and the $\eta_{\sigma}$ stands for the two-component spinors.

With these definitions, it can be easily shown that the application of the variation
principle to Eq. (2.3) leads to a coupled integro-differential equations in 
$P_{n\kappa}(r)$ and $Q_{n\kappa}(r)$. Therefore, to obtain the numerical 
wave-functions, we have to solve these two coupled integro-differential equations. 
Alternatively, a pseudo-eigenvalue equation (Hartree-Fock-Roothan) \cite{Roothan}
can be obtained by using an analytic expansion-type wave-functions and minimizing 
the energy functionals $E$ with respect to the expansion coefficients.

It has been found that the numerical wave-functions have more accurate asymptotic
behavior than the analytical ones, though both provide total energies of comparable
accuracy. The accuracy of the total energy and wave-function obtained through 
the Dirac-Fock-Roothan equation (FBSE method) can {\it in principle} be enhanced to 
any degree by increasing the number of basis functions, but {\it in reality} only 
a finite number of basis can be used because the computational time increases very 
rapidly with the increasing number of basis functions. Moreover, the use of large 
basis functions severely impedes the efficiency of the post-Dirac-Fock computations. 

In the present paper, we introduce a hybrid scheme to solve the DF equation
through the pseudo-eigenvalue approach where basis functions are defined on a grid
and one- and two-electron radial integrals are evaluated numerically as opposed
to the conventional relativistic Hartree-Fock-Roothan equations. Since, the basis 
functions are defined on a grid and the matrix elements appearing in the relativistic 
Hartree-Fock-Roothan equations are evaluated numerically, this scheme can be regarded 
as a combination of numerical and analytical approach to the solution of DF-SCF
equation. Here, like the traditional analytical basis set expansion approach, the large 
and small components of the radial wave-functions are expressed as linear combination 
of basis functions, i.e.,
\begin{equation}
P_{n\kappa}(r)=\sum_p C^L_{\kappa p}g^L_{\kappa p}(r)
\end{equation}
and
\begin{equation}
Q_{n\kappa}(r)=\sum_p C^S_{\kappa p}g^S_{\kappa p}(r)
\end{equation}
where the summation index $p$ runs over the number of basis functions $N$, 
$g^L_{\kappa p}(r)$ and $g^S_{\kappa p}(r)$ are basis functions belonging
to the large and small components, respectively, and $C^L_{\kappa p}$ and
$C^S_{\kappa p}$ are the corresponding expansion coefficients. Though, any
basis functions can be used, we have chosen Gaussian-type of orbitals (GTOs) 
that has the following form for the large component,
\begin{equation}
g^L_{\kappa p}(r)={\cal N}^L_{p}r^{n_{\kappa}}e^{-\alpha_pr^2}
\end{equation}
with
\begin{equation}
\alpha_p=\alpha_0\beta^{p-1}
\end{equation}
where $\alpha_0,\beta$ are user defined constants, $n_{\kappa}$ specifies the orbital 
symmetry (1 for $s$, 2 for $p$, etc.) and ${\cal N}^L_{p}$ is the normalization factor 
for the large component. 
The small component part of the basis function is obtained by imposing the
kinetic balance and has the following form
\begin{equation}
g^S_{\kappa p}(r)={\cal N}^S_{p}(\frac{d}{dr}+\frac{\kappa}{r})g^L_{\kappa p}(r)
\end{equation}
where 
\begin{equation}
{\cal N}^S_{p}=\sqrt{\frac{\alpha_p}{2n_{\kappa}-1}[4({\kappa}^2+\kappa+n_{\kappa})-1]}
\end{equation}
Using the above definitions, the Dirac-Fock-Roothan equation for closed shell system
can be cast into a pseudo-eigenvalue equation of the form
\begin{equation}
FC=SC\epsilon 
\end{equation}
where $F$ is the Fock matrix and $S$, $C$ and $\epsilon$ are overlap, eigenvector
and eigenvalue matrices, respectively.  This pseudo-eigenvalue equation is first 
transformed into an eigenvalue equation $F^{\prime}C^{\prime}=C^{\prime}\epsilon$, 
which on diagonalization produces the desired eigenvalues ($\epsilon$) and eigenvectors 
($C=S^{-1/2}C^{\prime}$). Since, the detailed derivation of relativistic 
Hartree-Fock-Roothan equation and its matrix elements (analytical as well as numerical 
form) has been discussed in details by several authors \cite{Kim,Grant}, we conclude 
this section by reiterating that the DF matrix elements appearing in the hybrid 
relativistic Hartree-Fock- Roothan equations are evaluated numerically to avoid the 
evaluation of complicated analytical expression of two-electron matrix elements and 
to improve the accuracy of the orbital properties. The present procedure also provides 
an easy route to implement $N_c$ ($N_c$ specifies the number of occupied orbitals) 
operations instead of $N^2$ ($N$ denotes the number of basis functions) for the evaluation
of the two-electron radial integrals that appear in DF-SCF equation. A brief outline
of the scheme is the following:

In the SCF procedure, the integrals and the matrices are evaluated over the members
of the basis set ${\{\phi_{\mu}}\}$ rather than over the members of the set of solutions
${\{\psi_i}\}$ because the atomic or molecular orbitals (solutions of SCF equations) are 
not known until the calculation is complete. Since these two sets of functions are 
related by
{\begin{equation}
\psi_i=\sum_{\mu =1}^N C_{\mu i}\phi_{\mu}
\end{equation}
the two-electron matrix element of $F$ (the Hartree-Fock potential term) in ${\{\phi}\}$ 
basis can be written as
\begin{equation}
U_{ij}=\sum_{c}\langle\phi_i\psi_c|\frac{1}{r_{12}}|\phi_j\psi_c\rangle\equiv
\sum_c\sum_{\mu}\sum_{\nu}C^*_{\mu c}C_{\nu c}\langle\phi_i\phi_{\mu}|\frac{1}{r_{12}}|
\phi_j\phi_{\nu}\rangle
\end{equation}
which involves a two-index tranformation. However, this two-index transformation process can
be easily avoided by evaluating the $U_{ij}$ matrix elements in a mixed basis i.e., in 
${\{\phi,\psi}\}$ basis. This is trivial,  because the occupied orbitals can be updated 
(like density matrix) during the SCF iteration and, therefore, the two-electron matrix 
element $\langle\phi_i\psi_c|\frac{1}{r_{12}}|\phi_j\psi_c\rangle$ can be directly computed 
at each iteration without invoking two two-index transformation.

\section{Theory}
\subsection{Overview of multi-reference MBPT method}
Multi-reference many-body perturbation theory (MR-MBPT) may be regarded as
a reformulation of the exact Schr{\"o}dinger equation into a small reference
space that is subspace of the full Hilbert space. This reduction is achieved
by first separating the atomic or molecular orbitals into three sets: the core
${\{c}\}$, valence ${\{v}\}$ and the excited orbitals ${\{e}\}$ and then by
introducing projection operators $P$ for the reference space (also called
valence or model space) and $Q$ for its orthogonal complement or virtual space,
\begin{equation}
P=\sum_{i=1}^d|\alpha\rangle\langle\alpha|
\end{equation}
and
\begin{equation}
Q=1-P=\sum_{m=d+1}^{\infty}|m\rangle\langle m|
\end{equation}
where the sets ${\{\alpha}\}$ and ${\{m}\}$ are, respectively, reference and
complementary space functions and $d$ is the dimensionality of the reference
space. With the aid of these two projectors, the exact N-electron time-independent
Schr{\"o}dinger equation can be transformed into the equation
\begin{equation}
H_{eff}|\Psi_{\lambda}^0\rangle=E|\Psi_{\lambda}^0\rangle
\end{equation}
involving the effective Hamiltonian $H_{eff}$,
\begin{equation}
H_{eff}=PHP+PHQ(E-QHQ)^{-1}QHP
\end{equation}
where $H$ is the exact Hamiltonian, $H_{eff}$ acts only on the reference space
spanned by ${\{\alpha}\}$ and produces the exact eigenvalues $E$ for the selected
states as given by the full-space Schr{\"o}dinger equation.

Certain approximations are necessary to solve Eq. (3.4) since the right hand
side involves the unknown eigenvalue $E$ and $Q$ space states, which, in principle
are of infinite dimension. The expansion of the denominator about the zeroth order 
eigenvalue transforms the Brillouin-Wigner type $H_{eff}$ [Eq. (3.4) to the
Rayleigh-Schr{\"o}dinger type effective Hamiltonian ($H_{eff}$)
\begin{equation}
H_{eff}=PHP+PHQ(E_0-H_0)^{-1}QHP+\cdots
\end{equation}

The exact Hamiltonian $H$ is partitioned into $H_0$ (the zeroth order Hamiltonian)
and $V$ (the perturbation), where the zeroth order Hamiltonian is taken to be diagonal
in $P$ and $Q$ subspaces, and may be written as a sum of diagonal one-electron operators
$h_0$ defined by
\begin{equation}
h_0=\sum_i \epsilon_i|i\rangle\langle i|
\end{equation}
where $i$ runs over all orbitals and $\epsilon_i$ is the $i$th orbital energy.
The partitioning of the orbitals must ensure a well-defined separation of the
orbital energies between core, valence and excited orbitals. Failure to meet
this requirement introduces numerical instabilities into the perturbative
computations. Although, in the above we have chosen the zeroth order energy
to be the "sum over orbitals", this choice is {\it in principle}, at our 
disposal, and {\it in practice}, it strongly affects the convergence
properties  of the perturbative expansions. There are two general categories 
known as the generalized M{\"o}ller-Plesset (MP) \cite{MP} and the generalized 
Epstein-Nesbet (EN) partitioning scheme \cite{Nesbet}. The generalized MP 
partitioning utilizes a "sum over orbitals" treatment, whereas the generalized EN
pursues a "sum over states" formulation in constructing the zeroth order Hamiltonian 
$H_0$. Different potentials may also be invoked to construct $H_0$ and a wide range 
of potentials have been chosen \cite{Kelly,Finley} with varying degrees of success. 
The Hartree-Fock potential is the most widely used potential for MBPT computations, 
because many terms automatically vanish for this particular choice. The present
second order MBPT computation employs MP partitioning where the zeroth order energy
is constructed from the Hartree-Fock potential.

\subsection{Relativistic many-body perturbation theory }
The relativistic Dirac-Coulomb Hamiltonian (presented in section II) for a 
many-electron system may also be partitioned into $H=H_0+V$, where
\begin{equation}
H_0=\sum_{i=1}^N[c\vec\alpha_i.\vec p_i+(\beta_i-1)mc^2]-\sum_{i=1}^N\frac{Z(r_i)e}{r_i}
\end{equation}
and 
\begin{equation}
V=\frac{1}{2}\sum_{i\ne j}\frac{e^2}{|\vec r_i-\vec r_j|}
\end{equation}
Here, we have introduced an r-dependent nuclear charge to account for the finite
size of the nucleus which can provide sizable effects for $s_{1/2}$ and $p_{1/2}$ 
states. While pursuing a many-body perturbative calculations, it is advantageous 
(from numerical point of view) to express the zeroth order Hamiltonian as
\begin{equation}
H_0=\sum_{i=1}^N[c\vec\alpha_i.\vec p_i+(\beta_i-1)mc^2]-\sum_{i=1}^N\frac{Z(r_i)e}{r_i}
+\sum_{i=1}^NU(r_i)
\end{equation}
which subsequently redefines the perturbation $V$ 
\begin{equation}
V=\frac{1}{2}\sum_{i\ne j}\frac{e^2}{|\vec r_i-\vec r_j|}-\sum_{i=1}^NU(r_i)
\end{equation}
where the single-particle operator $U(r_i)$ is introduced to account the effective
(or average) potential experienced by an electron due to the presence of other 
electrons and is known as the Hartree-Fock potential. The Schr{\"o}dinger equation 
of the zeroth order Hamiltonian $H_0$  provides a set of orbitals that are first 
partitioned into core, valence and excited orbitals and then two projectors $P$ 
and $Q$ are introduced to cast the N-electron Schr{\"o}dinger equation into an 
effective Hamiltonian equation [Eq. (3.5)]. Finally, the effective Hamiltonian matrix 
is diagonalized to obtained the desired eigenvalues. [ Note that while carrying out 
relativistic MBPT calculations, negative energy states are excluded from sum over 
intermediate states ($|m\rangle$) to avoid {\it continuum dissolution}.]

The theoretical ionization process is usually described as $M\rightarrow$ 
$M^+$ +e. However, the ionization process may also be represented as 
$M^++e\rightarrow M$.  That is to say, ionization potential can be computed 
either by estimating the energy required to remove an electron (IP) from the 
neutral atomic or molecular system or evaluating the energy released during 
the electron attachment process (EA) to its positively charged counterpart. 
Though, {\it in principle} the computed energies will be the same, but 
{\it in practice} the theoretical treatment these two processess are not equally
convenient. For alkali-metal atoms or systems with one electron in the outermost
shell, it is convenient to estimate the ionization potential by computing the
energy released due to the addition of an electron to its positively charged
species (a closed shell core). It is to be emphasized that although theoretically, 
the above two processes should provide identical numbers, but for a truncated many-body 
calculation they need not be the same, because the orbitals and their corresponding 
energies are not identical in these two situations. In the first case the core and 
virtual orbitals experiences the potential due to the valence electron (singly occupied 
orbital), but this potential is not present in the second case.

The second quantized representation of electron attachment process to a closed shell 
core is:
\begin{equation}
|\Psi^0_{\lambda}\rangle=a^{\dagger}_v|\Phi_0\rangle=
a^{\dagger}_v(\Pi_{c=1}^N a_c^{\dagger})|0\rangle
\end{equation}
Here $|0\rangle$ and $|\Phi_0\rangle$ represents the true and closed-shell vacuum
state, respectively.  Operators $a^{\dagger}_v$ and $a^{\dagger}_c$ denote the 
valence and core creation operator, respectively. For Na-like system, the core 
orbitals are $1s_{1/2},2s_{1/2}, 2p_{1/2}$ and $2p_{3/2}$, and the valence orbitals 
will be $3s_{1/2}$,$3p_{1/2}$ and $3p_{3/2}$. For convenience, we introduce the 
notation $\alpha,\beta,\gamma,\cdots $ for core orbitals, $p,q,r,\cdots$ for excited 
orbitals, $u,v,w,\cdots$ for valence and $m,n$ for valence and/or excited orbitals. 

The second quantized representation of the zeroth order Hamiltonian $H_0$ and the 
perturbation $V$ are
\begin{equation}
H_0=\sum_{i,j} h_{ij}a^{\dagger}_ia_j+U_{ij}a^{\dagger}_ia_j
\end{equation}
and
\begin{equation}
V=\frac{1}{2}\sum_{ijkl} g_{ijkl}a^{\dagger}_ia^{\dagger}_ja_la_k-U_{ij}a^{\dagger}_ia_j
\end{equation}
where 
\begin{equation}
h_{ij}=\int d^3r \psi_i(\vec r)^{\dagger}[c\vec\alpha_i.\vec p_i+(\beta_i-1)mc^2-
\frac{Z(r_i)e}{r_i}] \psi_j(\vec r)
\end{equation}
\begin{equation}
U_{ij}=\int d^3r \psi_i(\vec r)^{\dagger} U(\vec r) \psi_j(\vec r)
\end{equation}
and
\begin{equation}
g_{ijkl}=e^2\int\int \frac{d^3r_1d^3r_2}{|\vec r_1-\vec r_2|}
\psi_i(\vec r_1)^{\dagger} \psi_j(\vec r_2)^{\dagger}\psi_k(\vec r_1)\psi_l(\vec r_2)
\end{equation}

Using these definitions, the second order effective Hamiltonian matrix [Eq. (3.5)] for 
electron attachment (EA) and detachment (IP) process can be expressed in terms of
single particle orbital (for Hartree-Fock potential) as
\begin{equation}
EA^{(2)}=\epsilon_v+\sum_{\alpha,m,n}\frac{g_{\alpha vmn}\bar g_{mn\alpha v}}
{\epsilon_v+\epsilon_{\alpha}-\epsilon_m-\epsilon_n}-
\sum_{\alpha,\beta,m}\frac{g_{\alpha\beta vm}\bar g_{vm\alpha\beta}}
{\epsilon_{\alpha}+\epsilon_{\beta}-\epsilon_v-\epsilon_m}
\end{equation}
and
\begin{equation}
IP^{(2)}=-\epsilon_{\alpha}-\sum_{\beta\gamma,p}\frac{g_{\beta\gamma\alpha p}
\bar g_{\alpha p\beta\gamma}} {\epsilon_{\beta}+\epsilon_{\gamma}-\epsilon_{\alpha}-\epsilon_p}+
\sum_{\beta,p,q}\frac{g_{\beta\alpha pq}\bar g_{pq\beta\alpha}}
{\epsilon_{\alpha}+\epsilon_{\beta}-\epsilon_p-\epsilon_q}
\end{equation}
where $\epsilon$'s are the single particle orbital energies and $\bar g_{ijkl}$ represents
\begin{equation}
\bar g_{ijkl}=g_{ijkl}-g_{ijlk}
\end{equation}
While the first term of Eqs. (3.17)-(3.18) accounts for the PHP of Eq. (3.5),
the second and third terms of Eqs. (3.17)-(3.18) represents the second term
of Eq. (3.5). The first terms of Eqs. (3.17)-(3.18) are generally called the
Koopmans' EA/IP value. The second and third terms of Eqs. (3.17)-(3.18) are
the correlation and relaxation contribution to the second order EA/IP,
respectively.

The problems of continuum dissolution first occurs at second order because of
the appearance of energy denominator. Unless the restriction of summation over 
only positive energy states is in place, this could lead to a vanishing energy 
denominator.

\section{Results}

\subsection{Ionization potentials of neutral alkali metal atoms and group IIIA elements.}

We present the ionization potentials (IP) of alkali-metal atoms computed through second 
order MBPT in Table I and compare with experiments \cite{Moore} and with the second order 
perturbative calculations of Johnson {\it et al.} \cite{Johnson}. The only difference 
between these two theoretical calculations lie in the choice of basis functions (apart 
from the dimension of the basis function).  While Johnson {\it et al.} generate the basis 
through the B-spline method, we employ geometric-Gaussian function (with $\alpha_0=0.0052$ 
and $\beta=2.75$) to construct the atomic orbital basis.  The entire computation is 
performed with a basis that ranges from $20s$$15p$$15d$$15f$ (for Lithium) to 
$28s$$24p$$20d$$16f$$10g$ (for Francium). 

Table I clearly demonstrates that the accuracy in the ionization potential estimated 
through Koopmans' theorem (KT) \cite{Koopman} of alkali-metal atoms decreases with 
increasing atomic number. For instance, the Koopman ionization potential for s$_{1/2}$ 
state starts off with an accuracy of 1\% for 2s$_{1/2}$ state of Lithium and finally ends 
up with 12\% for 7s$_{1/2}$ state of Francium. We also observe similar trends for p$_{1/2}$ 
and p$_{3/2}$ states, where separation between p$_{1/2}$ and p$_{3/2}$ states increases 
(degenerate in Lithium) with increasing atomic number. 

Apart from Sodium, inclusion of the second order MBPT terms (relaxation and correlation 
contribution) significantly improves the agreement with the experiment, especially for the 
heavy alkalis. While the accuracy of our computed ionization potential for lighter atoms 
is similar to that of Johnson {\it et al.} \cite{Johnson}, the accuracy in the 
estimated IP for heavy atoms (Cesium and Francium) is better than theirs. In 
particular, our computed IP values for Cesium are comparable to CCSD (coupled cluster 
calculation with singles and doubles) of Eliav {\it et al.}\cite{Eliav}. This small but 
non-negligible difference in computed IP for Cesium and Francium between our results and 
that of Johnson {\it et al.} clearly is a basis set effect. However, it should be noted 
that while pursuing higher order MBPT calculations, the use of such a large basis will be 
highly computer intensive unless some deep-lying core and high-lying virtual orbitals are 
discarded from the calculations.

Table II compares the ionization potentials of group IIIA elements computed through 
second order MBPT with the experiments \cite{Moore}. We found several interesting features
for this series. First of all, unlike the alkali-metal atoms the Koopmans' IP values 
do not change appreciably down the series. Secondly, the second order MBPT provides
less accurate IP value for these elements compared the alkali-metal atoms. The 
deviation in computed IP values for these elements is quite expected because the 
non-dynamical correlation effects are quite large for these elements due to 
the quasi-degeneracy of the highest lying occupied orbitals. For example, the $2s_{1/2},2p_{1/2}$ 
and $2p_{3/2}$ orbitals of Boron are quasi-degenerate and hence, a MR-MBPT (multi-reference
many-body perturbation theory) treatment is absolutely necessary for this system to improve 
the accuracy and low order perturbative convergence rate.

Generally, the theoretical treatment of the electron attachment process is difficult 
because the correlation and relaxation effects tend to cancel each other (See Fig. 1) 
and the success of the theoretical treatment depends upon the relative importance of
these two effects.  For alkali-metal atoms, the relaxation effect is small compared to 
the correlation effect (especially for heavy alkalis) and, hence, they don't cancel 
each other. Fig. 1 also illustrates that while the contribution from the relaxation 
part is small and roughly the same for all the alkalis beyond Sodium, the correlation 
contribution steadily increases with the increasing atomic number. This pattern, 
however, may change at higher order MBPT and ,in fact, it has also been observed by 
Johnson {\it et al.}\cite{Johnson}. It is also evident from the Table I that 
correlation and relaxation effects are important for inner orbitals which indicates 
that the contribution of correlation and relaxation term will be large for deep lying 
core orbitals. The precise estimation of correlation and relaxation effects for the deep
lying (or inner) core, therefore, requires higher order many-body effects and, hence, 
it is imperative that high order perturbative computations (like the coupled cluster 
method) is necessary for the accurate estimation of core ionization.

\subsection{Excitation energies of neutral alkali metal atoms and elements of group IIIA.}
The direct computation of hole-particle excitation energy involves the matrix elements 
($H_{\alpha p}^{\beta q}$) which through second order MBPT can be written as
\begin{equation}
H_{\alpha p}^{\beta q}=\langle\Phi_{\beta}^q|H_{eff}^{(2)}|\Phi_{\alpha}^p\rangle=
\langle\Phi_0| a_{\beta}^{\dagger} a_q H^{(2)}_{eff} a_p^{\dagger}a_{\alpha}|
\Phi_0\rangle
\end{equation}
Appropriate expansion of $H_{eff}^{(2)}$ yields
\begin{eqnarray}
H_{\alpha p}^{\beta q}&=&[\epsilon_p \delta_{pq}+\sum_{r,s,\gamma}\frac{g_{q\gamma rs}
{\bar g}_{rsp\alpha}}{\epsilon_p+\epsilon_{\gamma}-\epsilon_r-\epsilon_s}
-\sum_{r,\gamma,\delta}\frac{g_{\gamma\delta pr}
{\bar g}_{qr\gamma\delta}}{\epsilon_{\delta}+\epsilon_{\gamma}-\epsilon_r-\epsilon_q}]
\delta_{\alpha\beta}\nonumber \\
&+&[-\epsilon_{\alpha}\delta_{\alpha\beta}-\sum_{r,\gamma ,\delta}
\frac{g_{\gamma\delta\beta r}{\bar g}_{r\alpha\gamma\delta}}
{\epsilon_{\delta}+\epsilon_{\gamma}-\epsilon_r-\epsilon_{\alpha}}
+\sum_{r,s,\gamma}\frac{g_{\alpha\gamma rs}{\bar g}_{rs\alpha\beta\gamma}}
{\epsilon_{\beta}+\epsilon_{\gamma}-\epsilon_r-\epsilon_s}]\delta_{pq}\nonumber \\
&+&{\bar g}_{q\alpha\beta p}+\sum_{r,\gamma}\frac{{\bar g}_{\alpha\gamma pr}
{\bar g}_{qr\beta\gamma}}{\epsilon_{\beta}+\epsilon_{\gamma}-\epsilon_q-\epsilon_r}
+\sum_{r,\gamma}\frac{{\bar g}_{q\gamma\beta r}{\bar g}_{r\alpha\gamma p}}
{\epsilon_p+\epsilon_{\gamma}-\epsilon_r-\epsilon_{\alpha}}
\end{eqnarray}
Here, the first two terms of the right hand side of Eq. (4.2) corresponds to
the matrix elements for electron attachment and detachment processes and
the next two terms corresponds to two-body effective interaction for excitation
process. In Eq. (4.2) the last sum excludes $\gamma=\alpha$ and $p=r$.
However, the computation of excitation energies for alkali-metal atoms 
involving the highest singly occupied (at Dirac-Fock level) is rather simple,
because the effective two-body interactions do not appear. For example, the
$2s_{1/2}\rightarrow 2p_{1/2}$ transition process for Lithium atom can be 
expressed as
\begin{equation}
|\Phi_{2p_{1/2}}\rangle=|1s^22p_{1/2}\rangle=a_{2p_{1/2}}^{\dagger}a_{2s_{1/2}}|1s^22s\rangle=
a_{2p_{1/2}}^{\dagger}a_{2s_{1/2}}a_{2s_{1/2}}^{\dagger}|1s^2\rangle\equiv
a_{2p_{1/2}}^{\dagger}|1s^2\rangle 
\end{equation}
Therefore, the $2s_{1/2}\rightarrow 2p_{1/2}$ transition energy through second order
MBPT reduces to
\begin{eqnarray}
\Delta E_{2s_{1/2}\rightarrow 2p_{1/2}}&=&\langle\Phi_{2p_{1/2}}|H^{(2)}_{eff}
|\Phi_{2p_{1/2}}\rangle -\langle\Phi_{2s_{1/2}}|H^{(2)}_{eff}|\Phi_{2s_{1/2}}
\rangle \nonumber\\
&\equiv& \langle 1s^2|a_{2p_{1/2}}H^{(2)}_{eff}a_{2p_{1/2}}^{\dagger}|1s^2
\rangle-\langle 1s^2|a_{2s_{1/2}}H^{(2)}_{eff} a_{2s_{1/2}}^{\dagger}|1s^2
\rangle 
\end{eqnarray}
A careful analysis shows that the quantity on the right hand side of Eq. (4.4) is
nothing but the difference in ionization potential value (in terms of neutral Lithium
atom) or difference in electron affinity value (in terms of positively charged Lithium 
atom).  Therefore, once the valence ionization potentials are known for these 
alkali-metal atoms,  the excitation energies involving highest singly occupied orbital 
(at the Dirac-Fock level) can be easily evaluated by computing the difference in 
ionization potential value. 

Excitation energies and oscillator strengths computed through second order MBPT 
(using second order energy and unperturbed dipole matrix element) for alkali-metal 
atoms and group IIIA elements are compared with the experiment \cite{Weise} in Tables 
III and IV. These tables demonstrate that the second order MBPT estimates the 
$s_{1/2}\rightarrow p_{1/2}$ transition energies more accurately than 
$s_{1/2}\rightarrow p_{3/2}$ for alkali-metal atoms. Here, we also find  that the 
error in the estimation of $s_{1/2}\rightarrow s_{1/2}$ transition energies are 
less (on an average) compared to $s_{1/2}\rightarrow p_{1/2}$ and 
$s_{1/2}\rightarrow p_{3/2}$ for alkali-metal atoms. While the second order single 
reference MBPT provides an accurate estimate for the excited states of alkali-metal 
atoms, it yields somewhat inaccurate (compared to alkali-metal atoms) excited state 
energies for the group IIIA elements. This deviation in the estimation of the 
excitation energies for elements of group IIIA is not unexpected, since the 
highest occupied $s$ and $p$ orbitals are fairly close-lying for these elements, 
configurations like $ns^2np(J=1/2)$ and $np^3(J=1/2)$ interact strongly with 
eachother \cite{das84}. Therefore, these two configuration state functions (CSFs) 
should be included
in the reference space for an accurate description of the ground and excited 
states. In other words, a multi-reference MBPT treatment is necessary for an accurate 
description of the ground and excited states for these elements. Since our second order
single reference space MBPT for the ground and excited state energy computations 
do not treat the CSF $np^3$ as a reference space states, these CSFs act as an intruder 
states \cite{Schucan}, and, thereby affect the perturbative convergence. An extensive 
study of this problem is underway.
\section{Conclusion}
We have presented valence ionization potentials, excitation energies and oscillator 
strengths of alkali-metal atoms and group IIIA elements computed through single
reference (SR) second order MBPT where the single particle orbitals are generated by 
solving the Dirac-Fock Hamiltonian in a finite Gaussian basis. Since, the present procedure 
computes the one and two-electron radial integrals numerically by defining the atomic 
orbitals on a grid, it is easy to implement $\approx N_c$ dependence ($N_c$= No. of 
occupied orbitals) for the number of operations needed to evaluate the two-electron 
integrals that appears in Dirac-Fock self-consistent field equation. The numerical 
accuracy, achieved for single valence electron atoms promises that this hybrid 
method will be accurate for other many-electron atomic systems and with some 
modifications, can also be applied to molecular systems. 
\section{Acknowledgement}
The authors acknowledge the financial support of the Department of Atomic Energy 
(No 37/15/97/-R\&D. II/1603). We thank Angom Dilip for initiating work on the
generation of the Gaussian basis set. We also 
thank to Prof. Debashis Mukherjee, Indian Association for the Cultivation of Science, for
valuable discussions.  The services and computer (RS10000) facilities made available 
by the Indian Institute of Astrophysics are gratefully acknowledged.

\newpage
\begin{table}
\tighten
\caption{Second order ionization  potential (in a.u.) of alkali-metal atoms.}
\begin{tabular}{cccccccc}
 Atom     & Ionizing orbital   & \multicolumn{4}{c}{This Work} 
& Others\tablenotemark[1] & Experiment \tablenotemark[2] \\
          &                    &  KT 
&$\Delta$\tablenotemark[3]& Second order& abs. error (\%)   & \\
\hline    
Li        &  $2s_{1/2}$        &  0.19631&0.00162&  0.19793& 0.11  & 0.19797& 0.19814 \\ 
          &  $2p_{1/2}$        &  0.12862&0.00134&  0.12996& 0.22  & 0.13001& 0.13024 \\
          &  $2p_{3/2}$        &  0.12862&0.00134&  0.12996& 0.22  & 0.13001& 0.13024 \\
          &  $3s_{1/2}$        &  0.07370&0.00034&  0.07404& 0.18  & 0.07415& 0.07418 \\
&&&&&&\\
Na        &  $3s_{1/2}$        &  0.18204&0.00558&  0.18762& 0.65  & 0.18790& 0.18886 \\ 
          &  $3p_{1/2}$        &  0.10947&0.00168&  0.11115& 0.40  & 0.11123& 0.11160 \\
          &  $3p_{3/2}$        &  0.10939&0.00167&  0.11106& 0.41  & 0.11119& 0.11152 \\
          &  $4s_{1/2}$        &  0.07003&0.00120&  0.07123& 0.49  & 0.07141& 0.07158 \\
&&&&&&\\
K         &  $4s_{1/2}$        &  0.14751&0.01181&  0.15932& 0.13  & 0.15994& 0.15952 \\ 
          &  $4p_{1/2}$        &  0.09568&0.00436&  0.10004& 0.31  & 0.10033& 0.10035 \\
          &  $4p_{3/2}$        &  0.09547&0.00430&  0.09977& 0.11  & 0.10005& 0.10009 \\
          &  $5s_{1/2}$        &  0.06095&0.00268&  0.06363& 0.13  & 0.06395& 0.06371 \\
&&&&&&\\
Rb        &  $5s_{1/2}$        &  0.13939&0.01393&  0.15333& 0.12  & 0.15430& 0.15351 \\ 
          &  $5p_{1/2}$        &  0.09078&0.00497&  0.09575& 0.48  & 0.09626& 0.09619 \\
          &  $5p_{3/2}$        &  0.08995&0.00474&  0.09469& 0.44  & 0.09518& 0.09511 \\
          &  $6s_{1/2}$        &  0.05861&0.00315&  0.06176& 0.00  & 0.06216& 0.06177 \\
&&&&&&\\
Cs        &  $6s_{1/2}$        &  0.12753&0.01519&  0.14272& 0.27  & 0.14511& 0.14310 \\ 
          &  $6p_{1/2}$        &  0.08556&0.00642&  0.09198& 0.15  & 0.09253& 0.09212 \\
          &  $6p_{3/2}$        &  0.08374&0.00567&  0.08941& 0.23  & 0.08996& 0.08962 \\
          &  $7s_{1/2}$        &  0.05515&0.00341&  0.05856& 0.19  & 0.05939& 0.05867 \\
          &  $7p_{1/2}$        &  0.04177&0.00219&  0.04396& 0.001 &        & 0.04393 \\
          &  $7p_{3/2}$        &  0.04106&0.00202&  0.04308& 0.001 &        & 0.04310 \\
          &  $8s_{1/2}$        &  0.00419&0.00962&  0.01381&      &        &         \\
&&&&&&\\
Fr        &  $7s_{1/2}$        &  0.13184&0.01926&  0.15110& 0.96   & 0.15271& 0.14967 \\ 
          &  $7p_{1/2}$        &  0.08584&0.00788&  0.09327& 0.69   & 0.09431& 0.09392 \\
          &  $7p_{3/2}$        &  0.08041&0.00542&  0.08583& 0.46   & 0.08656& 0.08623 \\
          &  $8s_{1/2}$        &  0.05605&0.00411&  0.06016&       & 0.06074&         \\
\end{tabular}
\smallskip
\noindent
\tablenotetext[1]{Reference \cite{Sapirstein}}
\smallskip
\tablenotetext[2]{Reference \cite{Moore}}
\noindent
\tablenotetext[3]{Correlation and relaxation contribution to ionization potential.}
\noindent
\end{table}                                                  

\begin{table}
\tighten
\caption{Second order ionization  potential (in a.u.) of group IIIA elements.}
\begin{tabular}{cccccccc}
 Atom     & Ionizing orbital   & \multicolumn{4}{c}{This Work}&Experiment \tablenotemark[2]\\
          &                    &  KT 
&$\Delta$\tablenotemark[3]& Second order& abs. error (\%)    & \\
\hline    
B         &  $2p_{1/2}$        &  0.27587&0.02984&  0.30571    & 0.25   & 0.30494 \\
          &  $2p_{3/2}$        &  0.27579&0.02983&  0.30562    &        &         \\
          &  $3s_{1/2}$        &  0.11451&0.00596&  0.12047    &        &         \\
          &  $4s_{1/2}$        &  0.05172&0.00168&  0.05340    &        &         \\
&&&&&&\\
Al        &  $3p_{1/2}$        &  0.19522&0.02444&  0.21966    & 0.15   & 0.21998 \\
          &  $3p_{3/2}$        &  0.19472&0.02439&  0.21911    &        &         \\
          &  $4s_{1/2}$        &  0.09709&0.00646&  0.10355    &        &         \\
          &  $5s_{1/2}$        &  0.04563&0.00224&  0.04787    &        &         \\
&&&&&&\\
Ga        &  $4p_{1/2}$        &  0.19609&0.02569&  0.22178    & 0.59   & 0.22046 \\
          &  $4p_{3/2}$        &  0.19275&0.02576&  0.21851    &        &         \\
          &  $5s_{1/2}$        &  0.09996&0.00697&  0.10693    &        &         \\
          &  $6s_{1/2}$        &  0.04670&0.00231&  0.04901    &        &         \\
&&&&&&\\
In        &  $5p_{1/2}$        &  0.18881&0.02666&  0.21541    & 1.30   & 0.21263 \\
          &  $5p_{3/2}$        &  0.17974&0.02544&  0.20518    &        &         \\
          &  $6s_{1/2}$        &  0.09383&0.00822&  0.10205    &        &         \\
          &  $7s_{1/2}$        &  0.04422&0.00307&  0.04729    &        &         \\
&&&&&&\\
Tl        &  $6p_{1/2}$        &  0.19863&0.02397&  0.22260    & 0.80    & 0.22446 \\
          &  $6p_{3/2}$        &  0.16602&0.02300&  0.18902    &        &         \\
          &  $7s_{1/2}$        &  0.09647&0.00789&  0.10436    &        &         \\
          &  $8s_{1/2}$        &  0.04526&0.00279&  0.04805    &        &         \\
\end{tabular}
\smallskip
\noindent
\tablenotetext[1]{Reference \cite{Sapirstein}}
\smallskip
\tablenotetext[2]{Reference \cite{Moore}}
\noindent
\tablenotetext[3]{Correlation and relaxation contribution to ionization potential.}
\noindent
\end{table}                                                  

\begin{table}
\tighten
\caption{Second order Excitation energies (in cm$^{-1}$) and oscillator 
strengths of alkali-metal atoms.}
\begin{tabular}{cccccccc}
 Atom     & Transition              &\multicolumn{3}{c}{This work}&\multicolumn{2}{c}{Experiment
\tablenotemark[1]}\\
          &                         &    Energy &abs. error(\%) &  Osc. Str.   &  Energy   & Osc. Str.   \\
\hline
Li        & $2s_{1/2}\rightarrow 2p_{1/2}$&    14915 &0.08 &   0.770      &  14903    & 0.753       \\
          & $2s_{1/2}\rightarrow 2p_{3/2}$&    14916 &0.08 &              &  14904    &             \\
          & $2s_{1/2}\rightarrow 3s_{1/2}$&    27188 &0.07 &              &  27206    &             \\
&&&&\\
Na        & $3s_{1/2}\rightarrow 3p_{1/2}$&    16783 &1.00 &   1.041      &  16956    & 0.982       \\
          & $3s_{1/2}\rightarrow 3p_{3/2}$&    16801 &1.00 &              &  16973    &             \\
          & $3s_{1/2}\rightarrow 4s_{1/2}$&    25544 &0.08 &              &  25739    &             \\
&&&&\\
K         & $4s_{1/2}\rightarrow 4p_{1/2}$&    13010 &0.19 &   1.231      &  12985    & 1.02        \\
          & $4s_{1/2}\rightarrow 4p_{3/2}$&    13068 &0.19 &              &  13043    &             \\
          & $4s_{1/2}\rightarrow 5s_{1/2}$&    21101 &0.36 &              &  21026    &             \\
&&&&\\
Rb        & $5s_{1/2}\rightarrow 5p_{1/2}$&    12636 &0.45 &   1.338      &  12579    &             \\
          & $5s_{1/2}\rightarrow 5p_{3/2}$&    12868 &0.40 &              &  12817    &             \\
          & $5s_{1/2}\rightarrow 6s_{1/2}$&    20097 &0.18 &              &  20134    &             \\
&&&&\\
Cs        & $6s_{1/2}\rightarrow 6p_{1/2}$&    11178 &0.40 &   1.414      &  11134    &             \\
          & $6s_{1/2}\rightarrow 6p_{3/2}$&    11700 &0.27 &              &  11732    &             \\
          & $6s_{1/2}\rightarrow 7s_{1/2}$&    18469 &0.36 &              &  18535    &             \\
&&&&\\
Fr        & $7s_{1/2}\rightarrow 7p_{1/2}$&    12692     &              &           &             \\
          & $7s_{1/2}\rightarrow 7p_{3/2}$&    14324     &              &           &             \\
          & $7s_{1/2}\rightarrow 8s_{1/2}$&    19957     &              &           &             \\
\end{tabular}
\noindent
\tablenotetext[1]{Reference \cite{Weise}}
\end{table}                                                  

\begin{table}
\tighten
\caption{Second order Excitation energies (in cm$^{-1}$) of group IIIA elements.}
\begin{tabular}{ccccccccc}
Atom     & Transition    &This work  & abs. error(\%)    &others\tablenotemark[1]&Experiment \tablenotemark[2]\\
\hline    
B         & $2p_{1/2}\rightarrow 3s_{1/2}$&    40656&1.54   &       & 40040       \\
          & $2p_{3/2}\rightarrow 3s_{1/2}$&    40635&1.53   &       & 40024       \\
          & $3s_{1/2}\rightarrow 4s_{1/2}$&    14720&1.66   &       & 14969       \\
&&&\\
Al        & $3p_{1/2}\rightarrow 4s_{1/2}$&    25483&0.51   &       &  25347      \\
          & $3p_{3/2}\rightarrow 4s_{1/2}$&    25362&0.51   &       &  25234      \\
          & $4s_{1/2}\rightarrow 5s_{1/2}$&    12220&0.99   &       &  12342      \\
&&&\\
Ga        & $4p_{1/2}\rightarrow 5s_{1/2}$&    25207&1.69   &       &  24789      \\
          & $4p_{3/2}\rightarrow 5s_{1/2}$&    24489&2.20   &       &  23962      \\
          & $5s_{1/2}\rightarrow 6s_{1/2}$&    12712&0.66   &       &  12796      \\
&&&\\
In        & $5p_{1/2}\rightarrow 6s_{1/2}$&    24880&2.08   &       &  24373      \\
          & $5p_{3/2}\rightarrow 6s_{1/2}$&    22634&2.14   &       &  22160      \\
          & $6s_{1/2}\rightarrow 7s_{1/2}$&    12018&0.75   &       &  11929      \\
&&&\\
Tl        & $6p_{1/2}\rightarrow 7s_{1/2}$&    25951&1.99   & 27048 &  26478      \\
          & $6p_{3/2}\rightarrow 7s_{1/2}$&    18581&0.56   & 19196 &  18685      \\
          & $7s_{1/2}\rightarrow 8s_{1/2}$&    12359&0.70   &       &  12268      \\
\end{tabular}
\noindent
\tablenotetext[1]{Reference \cite{Dzuba}}
\tablenotetext[2]{Reference \cite{Weise}}
\end{table}                                                  
\newpage
\begin{figure}[ht]
\unitlength0.7071mm
\begin{picture}(300,300)
\put(300,300){\makebox{\psfig{file=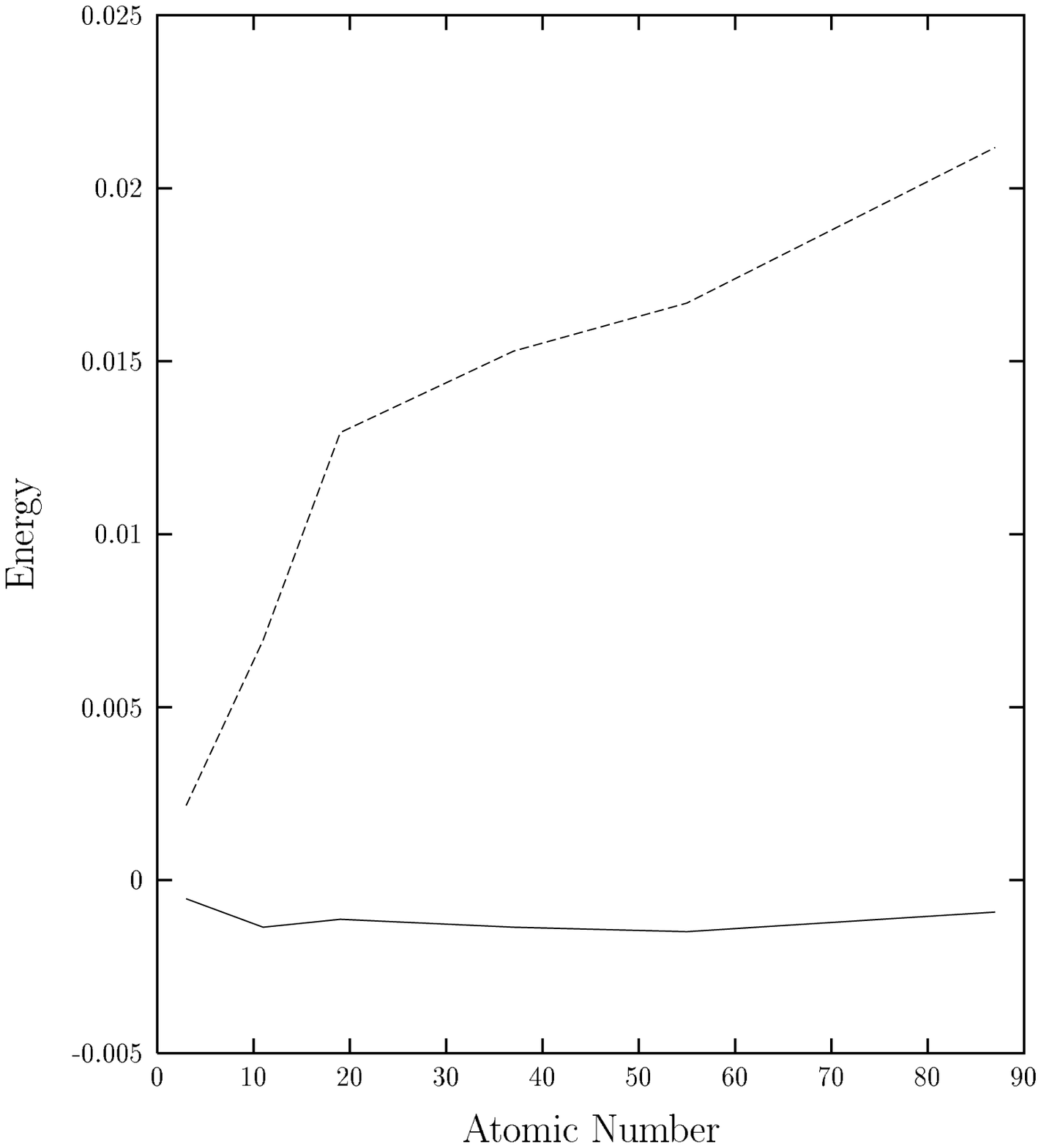,width=70mm}}}
\end{picture}                                                        
\caption{ Variation of correlation (dotted line) and relaxation energy (solid line)
as a function of atomic number for alkali atoms.}
\end{figure}                                          
\end{document}